# Measurement of Liquid Flow Rate among the Annular Flow in Vertical Tee Junction

**Shuo Huang [1,2], Tianyu Fu [3], Guanmin Zhang [3], Yu Sun[2,4],***

1. Laboratory of Signal and Image Processing, School of Biological Sciences and Medical Engineering, Southeast University, Nanjing, 210096, China.
2. International Laboratory for Children's Medical Imaging Research, School of Biological Sciences and Medical Engineering, Southeast University, Nanjing, 210096 China.
3. School of Energy and Power Engineering, Shandong University, Jinan, 250061 China.
4. Institute of Cancer and Genomic Science, University of Birmingham, Birmingham, B15 2TT, United Kingdom

\* Correspondence: Yu Sun (e-mail: sunyu@seu.edu.cn).
Orcid: Shuo Huang (0000-0002-4499-1702)

**Abstract:** Since the liquid flow rate of the annular flow is closely related to the heat exchange efficiency, it has great significance to measure the liquid flow rate of the annular flow in vertical tee junction. In order to acquire the liquid flow rate of the annular flow in vertical tee junction, a measurement method has been designed, which implements the digital subtraction method to measure the thickness of the liquid film under the visible light and to apply the image feature matching algorithm to obtain the liquid velocity field. Moreover, the accuracy of the liquid film velocity field as well as the spatial and temporal stability of the mass flow rate is tested by proposed algorithms in this study. Experimental results show that the measurement error of our method is approximately 5% in the lower section of the main pipe and the branch pipe, and lower than 15% in the upper section of the main pipe. Therefore, this method has a high accuracy in comparison with other measurement approaches. Our method can be applied to measure and analyse the shape and property of the annular flow in the vertical tee junction.

**Keywords:** Annular flow; Vertical tee junction; Liquid flow rate; Digital subtraction; Accuracy validation

## 1. Introduction

The annular flow in the vertical tee junction is widely existed in heat exchange equipment such as refrigeration equipment and chemical engineering equipment [1]. The annular flow pattern is characterized by the flow of a thin and wavy liquid film along the inner wall, with a core of gas flowing in the centre [2, 3]. It has greater significance to measure the liquid flow rate of the annular flow in vertical tee junction, since the annular flow is a commonly existed flow pattern, and the liquid flow rate of the annular flow is closely related to the thermal substance exchange efficiency [4-6]. For example, the annular flow is the major flow pattern of the liquefied natural gas in large cold boxes, the liquid film thickness of which has a direct influence on the heat transfer efficiency.

In recent years, a lot of methods have been designed to measure the flow rate of two-phase flow. In accordance with whether separating the two-phase flow into single-phase in the measurement, these methods can be divided into separation and non-separation methods [7, 8]. Moreover, the non-separation methods can be further divided into contact and non-contact approaches.

The separation method is the method which firstly employs a separation device to separate the gas and liquid and then measure them separately [9, 10]. One simple approach contains a gas-liquid separator, a gas flowrate meter and a liquid flowrate meter. The flowrate of gas and liquid can be obtained separately after gas-liquid separation [11]. The separation method has high measurement stability and accuracy, which can also overcome the influence of flow instability and flow pattern changes. The extracting and separating method [8] designed by D. Wang et al not only has the advantages mentioned above, but also effectively reduces the instrument volume. It is an efficient approach of separation measurement.

A large number of non-separation approaches have been proposed in recent years. Some representative methods are listed below:

**a. The approaches using a single flowmeter**, such as the laser Doppler velocimeter [12], which can simultaneously measure the flow rate of both gas and liquid. It has been widely used in gas-liquid two-phase flow for many years. This device has the advantages of non-contact, high spatial resolution, and fast dynamic response. However, it is also expensive and complicated.

**b. The measurement method which employs combined flowmeters.** For example, the throttling element - volumetric flowmeter, double slotted plate, throttling element - gear flowmeter, target flowmeter - turbine flowmeter, Venturi tube - turbine flowmeter, and orifice - Venturi tube [13-15]. D. Wang et al proposed an efficient approach using the combination of a multihole probe and an orifice plate [15], which can obtain high accuracy measurement result on both the flowrate and the dryness of the two-phase flow.

**c. A combination of the conventional flowmeter and phase retention meter.** Some two-phase flow measurement devices measure the flow rate of two-phase flow by first determining the velocity of each phase and the phase holdup and then calculating the flowrate of each phase [14, 16]. The meter used to measure the speed includes the turbine flowmeter, electromagnetic flowmeter, and differential pressure flowmeter. Methods used to measure the phase holdup can be ray method, conductance method, and tomography method.

G. B. Zhang et al designed a combination meter using turbine flowmeter and flap - type shunt conductance sensor is another representative method [16]. It can calculate the flow rate of each phase by measuring the total gas-liquid flowrate and the water-cut rate. The average error of the total gas-liquid flowrate measured by this method is 7.9%. This method has been applied in gas-liquid two-phase flow measurement in gas wells in Daqing Oilfield, China. Z. Huang et al's flow measurement equipment using an electrical capacitance tomography system and a Venturi flowmeter [17] can obtain the flowrate of bubble flow, stratified flow and slug flow. The accuracy of the flowrate measurement on annular flow is 10% - 20%.

**d. Digital subtraction method.** With the development of optical technology, digital subtraction method has been developed to measure the liquid film thickness, especially in the falling film flow analysis [18-21]. This method can be used to realize the overall measurement of the liquid film thickness without affecting the flow field. Moreover, the digital subtraction method has a high accuracy and a good real-time performance. However, these methods also have some shortcomings. Usually the instruments are expensive. Moreover, it is difficult for these methods to obtain the flowrate of the air.

Dupont et al [10] acquired the time-resolved water film thickness mapping on a non-transparent wall by taking advantage of the diffuse reflection of the near-infrared light in a narrow wavelength band on the wall surface. Although the digital subtraction method can provide a wide view field without affecting the flow pattern, it has not been applied on the two-phase flow measurement in the tee junction since the tilt or rotation of tee junction can affect the extraction accuracy of liquid film's cross-section area. Furthermore, the wavelength of the light source needs to be strictly limited [10, 19], which greatly increases the difficulty of the experiment and the workload.

In addition, the capacitance tomography method [22], pulsation method [23], nuclear magnetic resonance method [24] radioisotope tracing method [25], ultrasonic method [26] and correlation method [27, 28] are some other methods that have been widely studied.

In this work, a measurement method based on the digital subtraction method has been proposed to acquire the liquid flow rate of the annular flow in the vertical tee junction. For this purpose, firstly a digital subtraction method is implemented to measure the thickness of the liquid film under the visible light field, next the perspective transformation method is designed to correct the deflection of the image, then the liquid velocity field is calculated via the image feature matching algorithm, finally the accuracy of the liquid film velocity field and the spatial and temporal stability of the mass flow rate are tested by proposed algorithms in this study.

Experimental results show that the measurement error of our method is approximately 5% in the lower section of the main pipe and the branch pipe, and lower than 15% in the upper section of the main pipe. Therefore, this method has a high accuracy in comparison with other measurement approaches including the method based on the combination instrument of turbine flowmeter and conductance sensor in Reference [16], and the method which employed the electrical capacitance tomography technique and a Venturi tube in Reference [17]. Our method can be applied to measure and analyse the shape and property of the annular flow in the vertical tee junction.

## 2. Method

The average flow rate of the upper section and the lower section of the main pipe (denoted by $Q_{MT}$ and $Q_{MB}$, respectively) as well as the branch pipe (denoted by $Q_{BR}$) is acquired in this study [29]. The location of each section is shown in Figure 1(d). According to the definition of the mass flow rate, the mass flow rate of the liquid film (denoted by $Q$) is defined as the mass of the liquid film which passes a certain cross-section in a time unit, namely:

$$Q = \rho \times \overline{S} \times \overline{v} \quad (1)$$

where $\rho$, $\overline{v}$ and $\overline{S}$ are the mass density, average flow velocity and the average cross-sectional vector area of the liquid film, respectively. According to Equation (1), in order to extract and calculate the liquid flow rate in each section of the tee junction, the average flow velocity and the cross-sectional vector area of water need to be extracted.

### 2.1. Experimental System

The experimental system is shown in Figure 1(a), which contains two parts: the fluid flow network and the image observation system. In the fluid flow network, a tee junction which is made of quartz glass with a thickness of 2.5 mm was used as observation section. The main pipe is 250 mm in length and has an outer diameter of 50 mm. The branch pipe which is located in the middle of the main pipe is 175 mm in length and has an outer diameter of 25 mm. For the image observation system, a Memrecam HX-6E CMOS high-speed camera was used to capture the colourful images, and a fill light with the power of 500 W, as the light source to produce the visible light. The relative position of the high-speed camera, tee junction and fill light is shown in Figure 1(b). The imaging setup of our high-speed camera delivers a frame speed of 2000 frames per second, a FOV of approximately 217.6 mm × 214.4 mm (1088 pixel × 1072 pixel) and a projected pixel size of approximately 0.2 mm/pixel when placed at 0.5 m from the tee junction. Under this situation, the ISO value of the high-speed camera when capturing the RGB images is 10000, which is enough to meet our needs. Moreover, brown dye is employed in our experiment to improve the light absorption of the water. A photo of the annular flow is shown in Figure 1(d), and the photo of its corresponding empty tube is shown in Figure 1(c). The operating condition of this data is shown in the first experiment of Table 1 in Chapter 3.

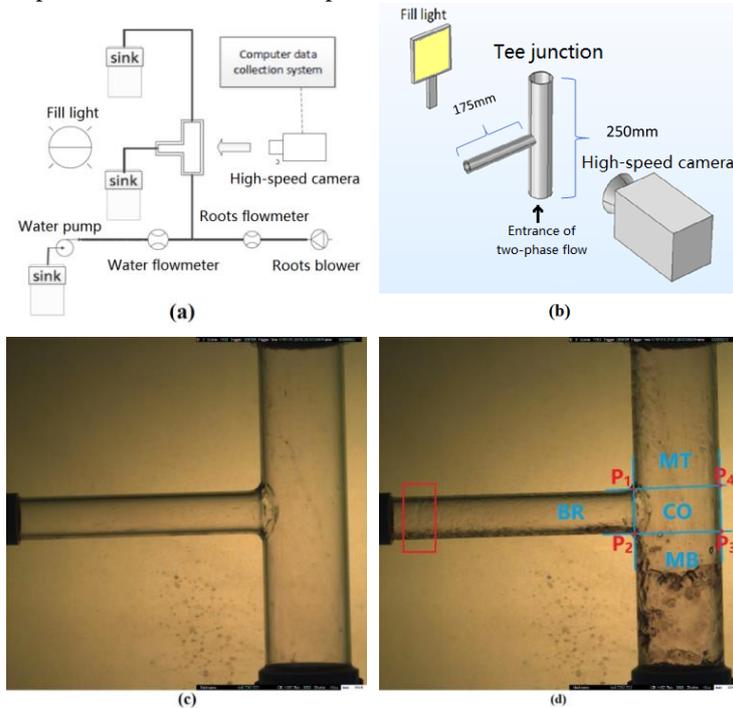

**Figure 1.** The experimental system. (a) the system chart of the experimental system. (b) the relative position of the high-speed camera, tee junction and fill light. (c) the reference image of Figure 1(d). (d) a photo of the annular flow.

In order to facilitate the calculation and reduce the amount of data, the original RGB colour photos are converted to the gray scale photos with the help of the following equation: $I = 0.299R + 0.587G + 0.114B$, where $I$ is the brightness of the gray scale photos and $R$, $G$ and $B$ represents the brightness of the red, green and blue channel of the correspondence RGB colour photo, respectively [30].

*2.2. The Calculation of Cross-sectional Area*

In this section, the digital subtraction method is applied to extract the thickness of the liquid film. After that, the cross-sectional area of the liquid film in each cross section of the tee junction can be calculated. However, the wavelength of the ambient light must be tightly constrained when using the traditional digital subtraction method. It greatly increases the difficulty of the experiment and limits the application of the method [10, 31]. Therefore, a method to analyse the thickness of liquid film under a visible light environment is proposed.

Since the fill light is a non-coherent light source, the brightness of the empty tube's photo (the reference luminance $I_{ref}$) can be defined as a superposition of the luminance of light under each wavelength contained in the filled light (defined as $I_{ref}\{0\}$, $I_{ref}\{1\}$, $I_{ref}\{2\}$, ..., respectively).

$$I_{ref} = I_{ref}\{0\} + I_{ref}\{1\} + I_{ref}\{2\} + \cdots = \sum_n I_{ref}\{n\}, n = 0,1,2, \quad (2)$$

According to Beer-Lambert's law, the relationship between the transmission of light under a certain wavelength (defined as $I\{n\}$) and the average thickness of the liquid film (defined as $\overline{d}$) on a certain pixel of image is:

$$\begin{cases} I\{n\} = I_{ref}\{n\} \times e^{-2k_n\overline{d}} \\ \overline{d} = \frac{d_1 + d_2}{2} \end{cases} \quad (3)$$

where $d_1$ and $d_2$ are the thickness of the liquid film at the given pixel and $k_n$ is the attenuation coefficient of water at the given wavelength.

Furthermore, the brightness of the corresponding pixel in the subsequent image, Figure 1(d), can be expressed as:

$$I = \sum_n \left( I_{ref}\{n\} \times e^{-2k_n\overline{d}} \right), \quad n = 0,1,2,\cdots \quad (4)$$

Equation (4) can be expanded with the Maclaurin formula:

$$I = \sum_n \left( I_{ref}\{n\} \times \sum_{m=0}^{+\infty} \left( \frac{(-2k_n\overline{d})^m}{m!} \right) \right), \quad n = 0,1,2,\cdots \quad (5)$$

Since the attenuation coefficient of our stained water at visible wavelengths is less than 1 cm$^{-1}$ due to our experiment, and according to the reference [32], the liquid film's thickness of the annular flow is also far less than 1 cm. Thus, we have $2k_n\overline{d} << 1$, and Equation (5) can be simplified as:

$$I = \sum_n \left( I_{ref}\{n\} \times \left(1 - 2k_n\overline{d}\right) \right) = I_{ref} - 2\overline{d} \sum_n \{(I_{ref}\{n\} \times k_n)\}, \quad n = 0,1,2,\cdots \quad (6)$$

Since the value of $\sum_n (I_{ref}\{n\} \times k_n)$ is independent of the film thickness, a proportional coefficient $k$ can be introduced to further simplify Equation (6), that is:

$$I = \sum_n \left( I_{ref}\{n\} \times \left(1 - 2k_n\overline{d}\right) \right) = I_{ref} - 2k\overline{d}I_{ref} \quad (7)$$

where $\sum_n (I_{ref}\{n\} \times k_n) = k \times I_{ref}$.

The average thickness of the liquid film can be calculated using the following equation to further reduce the error:

$$\begin{cases} I = I_{ref} \times \sum_{m=0}^{+\infty} \left( \frac{(-2k\overline{d})^m}{m!} \right) = I_{ref} \times e^{-2k\overline{d}} \\ \overline{d} = \frac{\ln\left(\frac{I}{I_{ref}}\right)}{-2k} \end{cases} \quad (8)$$

Since the background light may be converged by the liquid film in some regions due to the lens phenomenon, the brightness of these regions in Figure 1(d) is brighter than their corresponding regions in Figure 1(c). The obtained average thickness in these regions calculated through Equation (8) is less than zero. Directly setting the thickness of the liquid film at these locations to 0 will undoubtedly cause errors [10]. However, if we reverse the light path in these bright regions positions, the convergence of light can be regarded as a reverse of the light absorption process. Therefore, according to Equation (8), in these regions, we have:

$$\overline{d'} = \frac{\ln\left(\frac{I_{ref}}{I}\right)}{-2k} = -\overline{d}, \text{ where } \overline{d} < 0 \tag{9}$$

where $\overline{d'}$ is the average thickness of the liquid film at these bright regions. Thus, the average thickness of the liquid film can be calculated as:

$$\overline{d} = \left|\frac{\ln\left(\frac{I_{ref}}{I}\right)}{-2k}\right| \tag{10}$$

A method which is shown in Appendix A is employed to obtain the value of $k$ before the experiment, since the composition of light varies under different light sources. $k$ is determined to be 0.1cm$^{-1}$ under the operating condition of our experiment. Moreover, the liquid film was doped with colouring agent in our experiment to improve the accuracy.

The perspective transformation method is implemented to eliminate tilt and rotation of the image [19]. The quadrangle which is surrounded by the extension edge of the main pipe and the branch edge, quadrangle P$_1$P$_2$P$_3$P$_4$ in Figure 1(d), is employed to obtain the rotation matrix. Meanwhile, this quadrangle is also the boundary of each part of the tee junction. The average liquid film thickness map of the annular flow in Figure 1(d) is shown in Figure 2.

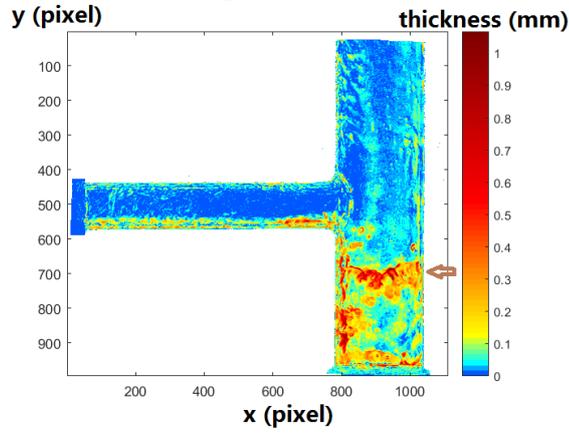

**Figure 2.** The average liquid film thickness map of the annular flow in Figure 1(d).

In order to calculate the cross-sectional area of the liquid film more accurately, the cubic polynomial fitting method are used to adjust the distribution of the data points on each cross-section and make the fitting points distributed evenly and more closely, which means that the distance of central angle between the neighbouring fitting points are equal to a small constant value θ. θ is set to be $\frac{\pi}{200}$ here, under which condition the corresponding arc length of the pipe shell is 0.3534 mm (1.7671 pixel) on the main pipe and 0.1571 mm (0.7854 pixel) on the branch pipe, which is close enough to acquire high accuracy calculation results. Moreover, most of the waves on the liquid film have larger wavelengths than 1.7671 pixel, so that this value of θ is enough to obtain high accuracy results. Equations 11(a) and 11(b) are used to calculate the average cross-sectional area of the liquid film:

$$S(q) = 2 \times \sum_{i=1}^{j}\left(\frac{1}{2}\theta \times \left(r^2 - \left(r - \overline{d_q(\iota\theta)}\right)^2\right)\right) = \theta \times \sum_{i=1}^{j}\left(\overline{d_q(\iota\theta)}\left(2r - \overline{d_q(\iota\theta)}\right)\right), \quad j = \frac{\pi}{\theta} \tag{11a}$$

$$\overline{S} = \frac{1}{n}\sum_{q=1}^{n} S(q) \tag{11b}$$

where r denotes the radios of the tee junction, $d_q(i)$ denotes the thickness distribution on the cross-section and n is the number of cross-section.

*2.3. The Calculation of Average Flow Velocity*

In this section, the feature-matching algorithm based on cross-correlation analysis [33, 34] is employed to acquire the displacement map of the liquid film. A detailed derivation of the feature-matching algorithm is given in Appendix B. The thickness distribution of the liquid film is employed as a feature in the calculation, as the shape of the liquid film remains stable within a short time interval. Moreover, it is also based on the fact that the unique distribution of the liquid film's depth in each window on one picture since the side length $m$ of the square windows is larger than the wavelength of waves of the liquid film, the positional relationship between these waves is usually complex, and the feature of each wave (such as the wavelength, shape and peak value) differs from each other. Due to the reasons above, the fluctuation of waves cannot lead to the highly matched data. When the data in two windows are matched, which means that the liquid film's depth distribution in these windows are almost unchanged, the only possible explanation is that the liquid film in these windows moves integrally. Therefore, our feature-matching method can be used to extract the integral displacement of the liquid film in each square window and calculate the integral flow velocity of liquid film in these square windows.

The average flow velocity of liquid film in a square window $\mathbf{W_n}$ with side lengths $m$ and starting point $(x_1, y_1)$ in figure $D_1$ can be calculated by the following equation:

$$\mathbf{v}(x_1, y_1) = \left(\frac{x_1-x_2}{\Delta t}, \frac{y_1-y_2}{\Delta t}\right) \qquad (12)$$

where $(x_2, y_2)$ denotes the starting point of the matched region $\mathbf{w_n}$ in figure $D_2$.

The liquid film velocity field with the side lengths of 32 pixels and 64 pixels are shown in Figures 3(a1) and (b1), respectively. Moreover, in order to facilitate the comparison, the detail information of the connection part between the main pipe and branch pipe has been enlarged and shown in Figures 3(a2) and (b2), respectively, since the liquid film velocity field in this part is more complexity than other parts. The direction and length of the arrows in Figure 3 reflect the orientation and magnitude of the liquid film's velocity in each square window, respectively. According to Figures 3, the velocity field with 64 pixels reflects more clearly the overall motion characteristic of the liquid film. Therefore, the average flow velocity is obtained under the side lengths of 64 pixels in our method.

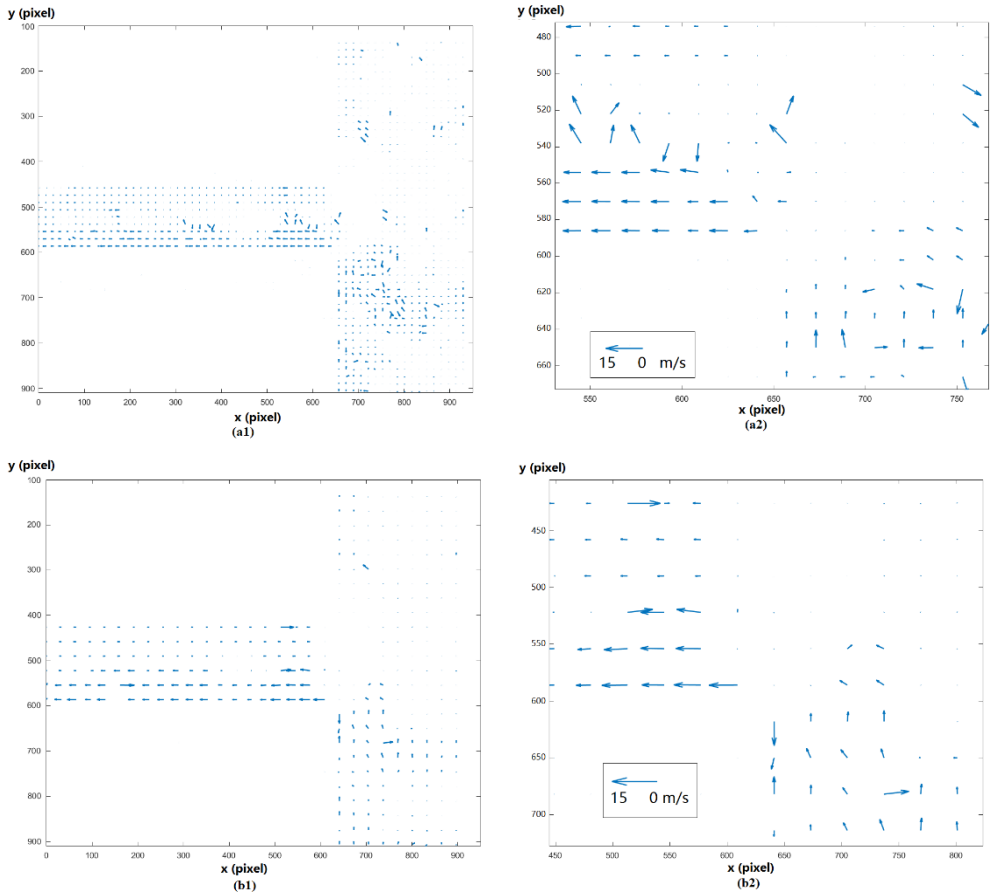

**Figure 3.** The liquid film's velocity field under different side lengths of the square windows, m. (a1) m = 32 pixels; (a2) the detail of Figure 3(a1); (b1) m = 64 pixels; (b2) the detail of Figure 3(b1). Δt = 0.0035 s.

*2.4. The Calculation of Liquid Flow Rate*

In this part, the average flow rate of the upper section and the lower section of the main pipe, as well as the branch pipe can be obtained by Equation (1). However, since the flow rate of the upper section is too small to be measured accurately, Equation (13) is used to get the average flow rate of the upper section:

$$Q_{MT} = Q_{MB} - Q_{BR} \qquad (13)$$

Furthermore, the proposed method is verified through measuring the true value of the average flow rate in each section of the tee junction and calculating the percentage error $\varepsilon$ between measured values and true values

The flow chart of the liquid flow rate calculation of the annular flow in the tee junction is shown in Figure 4.

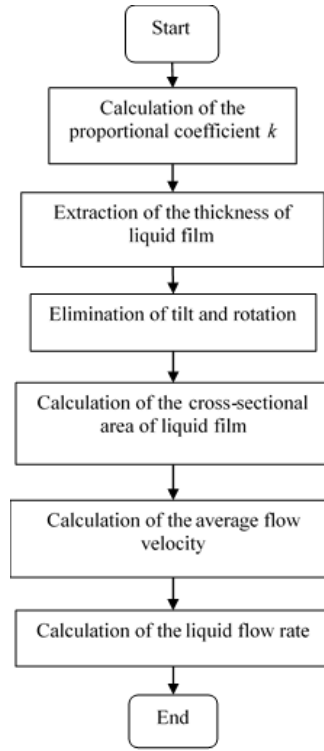

**Figure 4.** Flow chart of the liquid flow rate calculation of the annular flow in the tee junction

## 3. Results

Experiment under three groups of experimental conditions is designed and implemented to verify our method. The inlet air flow rate ranges from 1 to 2.5 $m^3 \times min^{-1}$, and inlet liquid flow rate is between 100 and 300 $mL \times min^{-1}$. The experimental conditions are shown in Table 1.

Table 1. The experimental conditions

| No. | inlet air flow rate ($m^3 \times h^{-1}$) | inlet air pressure (kPa) | inlet air temperature (°C) | inlet liquid flow rate $q_{MB}$ ($g \times s^{-1}$) |
|---|---|---|---|---|
| 1 | 89.59 | 111.74 | 21.27 | 4.14 |
| 2 | 153.64 | 117.66 | 27.44 | 4.43 |
| 3 | 125.07 | 115.03 | 24.25 | 4.77 |
| 4 | 131.20 | 115.32 | 26.43 | 1.19 |
| 5 | 130.47 | 115.63 | 26.35 | 2.35 |
| 6 | 129.36 | 115.94 | 26.76 | 3.52 |
| 7 | 129.77 | 116.42 | 26.56 | 5.90 |
| 8 | 128.93 | 116.86 | 27.35 | 7.78 |
| 9 | 127.80 | 117.25 | 27.63 | 9.02 |

In order to reduce the experimental error, the regions that have stabilized cross-sectional area of liquid film are applied to measure the average cross-sectional area of each section of the tee junction in our experiments. In addition, according to Equation (1), the liquid flow-rates in these regions are also stabilized. The average cross-sectional area of sections MT, MB and BR on the tee junction is shown in Table 2.

Table 2. The average cross-sectional area of sections MT, MB, and BR on the tee junction.

| No. | $\bar{S}_{MB}$ (mm²) | $\bar{S}_{BR}$ (mm²) | $\bar{S}_{MT}$ (mm²) |
|---|---|---|---|
| 1 | 7.9862 | 0.7677 | 2.2014 |
| 2 | 4.2776 | 1.0077 | 2.1897 |
| 3 | 5.3097 | 1.0141 | 1.9454 |
| 4 | 3.7039 | 1.1373 | 1.4279 |
| 5 | 4.0856 | 1.3132 | 1.7106 |
| 6 | 7.0120 | 1.7279 | 2.2054 |
| 7 | 11.9459 | 2.3750 | 2.9123 |
| 8 | 14.1513 | 2.7960 | 4.0715 |
| 9 | 17.0777 | 3.9270 | 4.4532 |

The average flow velocity of the liquid film in each section of the tee junction is listed in Table 3. However, since the flow velocity of the upper section of main pipe is too slow to be accurately measured, the average flow velocity of liquid film in this section is calculated through the following equation: $\bar{v}_{MT} = \frac{Q_{MT}}{\rho \times \bar{S}_{MT}}$.

Table 3. The average liquid flow velocity in each section of the tee junction

| No. | t (s) | $\bar{v}_{MB}$ (mm×s⁻¹) | $\bar{v}_{BR}$ (mm×s⁻¹) | $\bar{v}_{MT}$ (mm×s⁻¹) |
|---|---|---|---|---|
| 1 | 0.0020 | 514.29 | 4685.46 | 231.66 |
| 2 | 0.0020 | 1085.69 | 3885.35 | 332.75 |
| 3 | 0.0015 | 857.14 | 3942.63 | 284.12 |
| 4 | 0.0030 | 311.11 | 962.82 | 40.18 |
| 5 | 0.0020 | 466.67 | 1693.23 | 107.86 |
| 6 | 0.0020 | 491.43 | 1792.83 | 157.85 |
| 7 | 0.0020 | 531.43 | 2324.04 | 207.17 |
| 8 | 0.0015 | 533.33 | 2390.44 | 212.12 |
| 9 | 0.0020 | 560.00 | 2231.08 | 180.12 |

Table III shows that $\bar{v}_{BR} > \bar{v}_{MB} > \bar{v}_{MT}$. It is due to the fact that the gravity effect of liquid film gravity is nullified by the support of the branch pipe. Furthermore, the smaller diameter of the branch pipe further increased the speed of liquid film in the branch pipe. It is clearly illustrated in Tables 2 and 3 that the liquid film in the upper section of main pipe has the lowest flow velocity and is also the thickest. This is because the weakening of air pressure in the main pipe caused by the streaming effect of branch pipe leads to a weaken thrust of the liquid film in the upper section of main pipe. Additionally, the weakening of air pressure also causes the wave around the junction of the main pipe and branch pipe, which is pointed by arrows in Figure 2.

The calculation results of average mass flow rate in each section of the tee junction is calculated by Equation (1) and listed in Table 4. The percentage errors $\varepsilon$ between the calculation results and their corresponding true value of liquid film's mass flow rate (denoted by $q_{MB}$, $q_{MT}$ and $q_{BR}$, respectively) is employed to verify our algorithm. However, it is difficult to acquire $q_{MB}$, $q_{MT}$ and $q_{BR}$ directly due to the low accuracy of the measurement approaches, just as shown in Section 4.3, and we should find a way to acquire the high accuracy result of the true value.

Table 4. The experimental results of the mass flow rate in each section of the tee junction.

| No. | $q_{MB}$ (g×s⁻¹) | $Q_{MB}$ (g×s⁻¹) | $\varepsilon_{MB}$ | $q_{BR}$ (g×s⁻¹) | $Q_{BR}$ (g×s⁻¹) | $\varepsilon_{BR}$ | $q_{MT}$ (g×s⁻¹) | $Q_{MT}$ (g×s⁻¹) | $\varepsilon_{MT}$ |
|---|---|---|---|---|---|---|---|---|---|
| 1 | 4.1402 | 4.1072 | -0.80% | 3.6125 | 3.5972 | -0.42% | 0.5277 | 0.5100 | -3.35% |
| 2 | 4.4333 | 4.6443 | +4.76% | 3.7833 | 3.9156 | +3.50% | 0.6500 | 0.7286 | +12.09% |

| | | | | | | | | |
|---|---|---|---|---|---|---|---|---|
| 3 | 4.7650 | 4.5511 | -4.49% | 4.1433 | 3.9984 | -3.50% | 0.6217 | 0.5527 | -11.09% |
| 4 | 1.1900 | 1.1523 | -3.17% | 1.1400 | 1.0950 | -3.95% | 0.0500 | 0.0574 | +14.74% |
| 5 | 2.2533 | 2.4080 | +2.32% | 2.1700 | 2.2235 | +2.47% | 0.1833 | 0.1845 | +0.64% |
| 6 | 3.5233 | 3.4459 | -2.20% | 3.1733 | 3.0978 | -2.38% | 0.3500 | 0.3481 | -0.53% |
| 7 | 5.8967 | 6.1230 | +3.84% | 5.2600 | 5.5197 | +4.94% | 0.6367 | 0.6033 | -5.24% |
| 8 | 7.7867 | 7.5474 | -3.07% | 7.0033 | 6.6837 | -4.56% | 0.7833 | 0.8637 | +10.25% |
| 9 | 9.0233 | 9.5635 | +5.99% | 8.2033 | 8.7614 | +6.80% | 0.8200 | 0.8020 | -2.18% |

An indirect measurement method is used to get the true value. According to the analysis of the temporal stability of the liquid film's liquid mass flow rate in Section 4.2, the mass flow rate of the liquid film has a high temporal stability, which means that the variation of liquid mass flow rate during a short measurement time is very small. Therefore, the true value of liquid film's mass flow rate can be obtained by measuring the weight of liquid flowing out of the tee junction from the main pipe and branch pipe in the minute of the shooting progress and calculating the average liquid flow rate in that minute. Furthermore, in order to strengthen the convincing of our true value, the average liquid flow rate from two minutes before to two minutes after the shooting time was also obtained. The analysis of these data shows that the fluctuation of liquid film's mass flow rate is about 1% in the branch pipe and no more than 3% in the upper section of the main pipe. Thus, the average mass flow rate in the minute of the shooting progress can be used as the true value $q_{MB}$, $q_{MT}$ and $q_{BR}$.

As shown in Table 4, the percentage errors of the liquid film's mass flow rate at the branch pipe and the lower section of the main pipe is approximately 5%, while the percentage errors at the upper section of the main pipe is approximately 15%. Considering that the mass flow rate at the upper section of main pipe is much smaller than the mass flow rate at the other two sections, the percentage errors at the upper section must be higher under the same measurement accuracy. Experimental results show that our algorithm achieve the high accuracy results of the average mass flow rate in each section of the tee junction

## 4. Discussion

*4.1. The Accuracy of the Liquid Film Velocity Field*

The accuracy of the obtained liquid film velocity field needs to be confirmed, since the liquid film at different locations has different velocity, and the thickness of boundary layer [35] also affects the accuracy of calculated results.

In order to get the accuracy of the calculated velocity, a novel parameter $\Phi_n$ is employed in this work, which quantitatively describes the percentage of water in two matched regions whose velocity is equal to the calculated value of these regions. The detailed derivation of $\Phi_n$ is shown in Appendix C. The distribution map of $\Phi_n$ among the data in Figure 1 (d) with the side lengths of 64 and 32 pixels are obtained, as shown in Figure 5.

According to the $\Phi_n$ distribution maps in Figure 5 (a1) (b1), the $\Phi_n$ on the main pipe is higher than that of the branch pipe. The minimum value of $\Phi_n$ appears near the junction of the branch pipe and the main pipe. This is due to the fact that the liquid film changes from a semi-annular flow to the annular flow at this position, the volume of the liquid film in the corresponding windows changes significantly. Figure 5 (a2) - (b3) shows that $\Phi_n$ is generally large when the side length of square window, m, is 64 and 32 pixels.

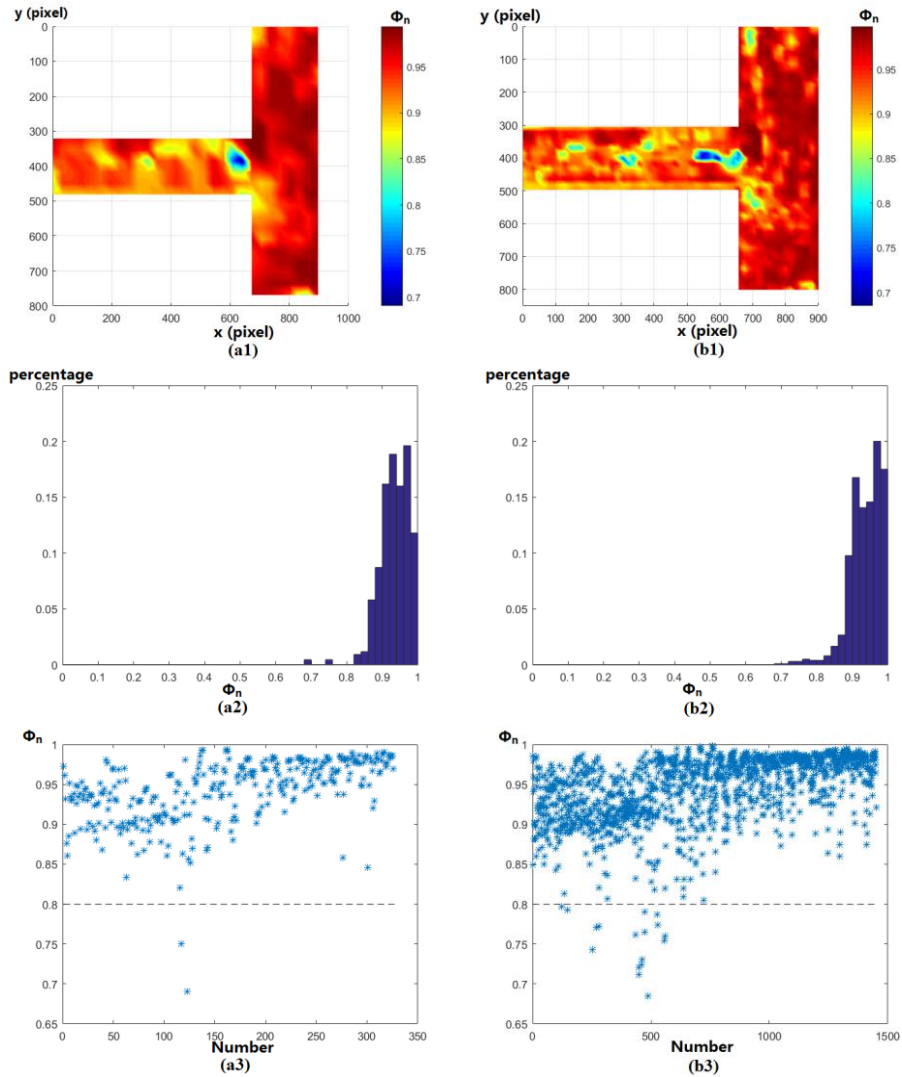

**Figure 5.** $\Phi_n$ distribution maps ((a1), (b1)), distribution histograms ((a2), (b2)), and the value of $\Phi_n$ in each window ((a3), (b3)) of experiment No. 1 when $m$ = 64 and 32 pixels, respectively.

Calculation results show that the percentage of the windows whose $\Phi_n$ is larger than 80% is 99.07% and 97.86% when $m$ is equal to 64 and 32 pixels, respectively. These results indicate that the gap between the calculated results and actual results of the liquid film's velocity on the experimental data in Figure 1(d) is narrow, which also means that the transition layer is thin and can be ignored. Therefore, the calculated results $v$ can be considered as the average speed of liquid film in each window. In addition, a larger percentage when the side length of square window $m$ is 64 pixels demonstrates that the larger the employed window in the calculation is, the smaller effect of the velocity field's detail on the obtained $v$ will be, which also confirmed the calculation results shown in Figure 3.

**Table 5.** The accuracy of the velocity field

| No. | Percentage of windows whose $\Phi_n$>80% when $m$=32 pixels | Percentage of windows whose $\Phi_n$>80% when $m$=64 pixels |
| --- | --- | --- |
| 1 | 97.86% | 99.07% |
| 2 | 90.73% | 90.11% |
| 3 | 94.83% | 98.76% |
| 4 | 92.36% | 95.64% |
| 5 | 96.45% | 98.62% |
| 6 | 95.94% | 98.42% |
| 7 | 95.13% | 97.35% |
| 8 | 93.25% | 94.36% |
| 9 | 89.46% | 90.42% |

The accuracy of the velocity field obtained under different operating conditions can be verified using the method above. Table 5 shows the percentage of the windows whose $\Phi_n$ is larger than 80% for each experiment in Chapter 3. As shown in Table 5 and Figure 5, the accuracy of velocity field obtained by our method is high under experimental conditions in Table 1. Thus, the average liquid flow velocity in Table 3 can be used to calculate the liquid mass flow rate of these data.

*4.2. Stability on the Mass Flow Rate of Liquid Film*

The stability of mass flow rate is of great significance when calculating the mass flow rate accurately. The spatial and temporal stability of the mass flow rate is used to evaluate the stability of the flow rate. The spatial stability is defined by the variance of the cross-sectional area of the liquid film in each section of the tee junction, which reflects the spatial deformation regularities of the liquid film. A high spatial stability indicates that the velocity of liquid film tends to be stabilized. Therefore, the high spatial stability part of each section is employed in the calculation of mass flow rate to get high-precision results.

The temporal stability is the changing rate of the liquid film's mass flow rate in each section of the tee junction. Since the changes of mass flow rate will inevitably lead to the change of liquid film's volume in the lumen, the changing rate of liquid film's volume per unit length of the tee junction at time $t$ (denoted by $u(t)$) is applied to measure the temporal stability of the tee junction. It is obvious that a smaller $u(t)$ denotes a slower changing rate of the liquid film's mass flow rate, which also means a higher temporal stability. Moreover, with the help of $u(t)$, the temporal stability can also be measured quantitatively.

The equation to calculate the temporal stability is shown in Equations (14) and (15), where $L$ and $N$ are the tube length and pixel length of the selected section of the tee junction, respectively. $Vol(t_n)$ is the volume of liquid film in the selected section at time $t_n$, $L_i$ is the length of one pixel, $\rho = 1.0$ g×mL$^{-1}$ is the density of water.

$$u(t) = \frac{\rho \cdot d(Vol(t))/dt}{L \cdot dt} = \frac{\rho(Vol_{OUT} - Vol_{IN})}{L(t_2 - t_1)^2} = \frac{\left(\int_{t_1}^{t_2} Q_{L_0+L}(t)dt - \int_{t_1}^{t_2} Q_{L_0}(t)dt\right)/(t_2 - t_1)}{L(t_2 - t_1)} = \rho \frac{Vol(t_2) - Vol(t_1)}{L(t_2 - t_1)^2} \quad (14)$$

$$\begin{cases} Vol(t_n) = \sum_{i=1}^{N} S(i) \times L_i, \ L_i = \frac{L}{N} \\ L \geq \frac{\bar{v}(t_2 - t_1)}{L_i} \end{cases} \quad (15)$$

**Table 6.** The temporal stability of the liquid film's mass flow rate

| No. | $u_{MB,t}$ (g×(s$^{-2}$×m$^{-1}$)) | $u_{BR,t}$ (g×(s$^{-2}$×m$^{-1}$)) | $u_{MT,t}$ (g×(s$^{-2}$×m$^{-1}$)) |
|---|---|---|---|
| 1 | -33.5490×10$^{-3}$ | 1.5114×10$^{-3}$ | -10.3934×10$^{-3}$ |
| 2 | -24.6950×10$^{-3}$ | 3.6794×10$^{-3}$ | -10.4135×10$^{-3}$ |
| 3 | -4.6442×10$^{-3}$ | -8.6859×10$^{-3}$ | -21.8858×10$^{-3}$ |
| 4 | 12.5663×10$^{-3}$ | 6.2576×10$^{-3}$ | 9.5263×10$^{-3}$ |
| 5 | -15.2637×10$^{-3}$ | 5.8094×10$^{-3}$ | 12.5046×10$^{-3}$ |
| 6 | 6.3592×10$^{-3}$ | -7.0982×10$^{-3}$ | -12.9538×10$^{-3}$ |
| 7 | 45.9615×10$^{-3}$ | -12.5638×10$^{-3}$ | 15.9284×10$^{-3}$ |
| 8 | -26.3427×10$^{-3}$ | -9.3264×10$^{-3}$ | 16.0249×10$^{-3}$ |
| 9 | -35.2918×10$^{-3}$ | 11.3879×10$^{-3}$ | -15.9278×10$^{-3}$ |

Table 6 shows the temporal stability of the liquid film's mass flow rate at experiments in Chapter 3. It can be found that at each experiment, the liquid flow's flow rate in the tee junction has a strong temporal stability. This result also demonstrates the accuracy of the true value of the mass flow rate obtained by the weighing method in Table 4.

*4.3. The Accuracy of the Proposed Method*

Experimental results show that the measurement error of our method is around 5% at the lower part of the main pipe and the branch pipe, and around 15% at the upper part of the main pipe. In comparison with the existing models, this method has a high measurement accuracy [36-38]. Compared with other measurement approaches, our method presents a smaller relative difference. For instance, the method based on the combination instrument of turbine flowmeter and conductance sensor with petal type concentrating flow diverter, has an average error of 7.9% and has been employed to measure the gas-liquid two phase flow in gas wells of Daqing oil field in China [16]. Additionally, some researches [17, 38] shows that the method which employed the electrical capacitance tomography technique and a Venturi tube presents an underestimation of about 10% to 20% in the annular flow. The accuracy of the methods above is worse than our method.

It should be noticed that tee junction is more complex than the pipe applied in these study, thus the mass flow rate is much more difficult to be extracted. Furthermore, none of these methods can provide the distribution map about either the thickness or the velocity field of the liquid film.

**5. Conclusions**

In this study, a method for acquiring the liquid flow rate of the annular flow in the vertical tee junction has been proposed. The digital subtraction technique under the visible light field is applied to extract the thickness distribution map, and the cross-sectional area of the annular flow is obtained. Then, the measurement of the liquid velocity field is implemented by the feature-matching algorithm based on cross-correlation analysis. After that, the mass flow rate of liquid film in branch pipe and the upper and lower sections of main pipe under different experimental conditions is acquired. Furthermore, the spatial and temporal stability of the mass flow rate is employed to evaluate the stability of flow rate. Experimental results show that the measurement error of our method is approximately 5% in the lower section of the main pipe and the branch pipe, and lower than 15% in the upper section of the main pipe. Therefore, this method has a high accuracy in comparison with other measurement approaches.

Future work in this area could involve the measurement of the irregular flow pattern such as the slug flow; applying the method to the research of other common type of cell structures; and analysing the spatial and temporal variation tendency on different flow parameters of liquid film under certain working condition; as well as the variation of flow parameters in the tee junction under different inlet conditions [39].

**Acknowledgments**

The authors thank Prof. Siyuan He, Ping Zhou and Suiren Wan in Southeast University and Prof. Min Wei and Xueli Leng from Shandong University for providing the experimental instruments.

**Funding:**

This research was supported by the National Natural Science Foundation of China under grant numbers. 51576115, 61127002, 11572087 and 31300780, the Research Center for Learning Science of Southeast University under grant number 3207038391, the School of Biological Sciences and Medical Engineering of Southeast University under grant number 3207037434 and the Fundamental Research Funds for the Central Universities under grant number 2242016R30014.

**Conflict of interest**

The authors declare no conflict of interest.

**Appendix**

*Appendix A. The Calculation of k*

(1). Firstly, an empty tank is set at the position of tee junction. Shooting its reference image $I_{emp}$;
(2). Pouring out some water droplets on inner wall of the tank, and shooting its image $I_w$;

(3). Calculating **d$_{tem}$** with Equation (A.1), where (x, y) denotes the coordinate of each pixel;

$$d_{tem}(x,y) = \ln \frac{I_w(x,y)}{I_{emp}(x,y)} \quad (A.1)$$

(4). Select a water droplet whose diameter is less than 3 mm on image **I$_w$**. This water droplet is hemispherical since its surface tension is much larger than its gravity;

(5). After calculating the core position and radius of the water droplet [40], the thickness at each pixel can be obtained. Then *k* can be acquired with Equation (A.2), where *k(i)* and *d(i)* denote the *k* value and the thickness of the *i*-th pixel, respectively. *j* is the number of pixels contained in the water droplet.

$$\begin{cases} k(i) = \frac{d_{tem}(i)}{d(i)}, \; 1 \leq i \leq j \\ k = \frac{1}{j}\sum_{i=1}^{j} k(i) \end{cases} \quad (A.2)$$

*Appendix B. Image Feature Matching Algorithm Based on Cross Correlation Analysis*

According to the cross-correlation analysis algorithm, the correlation coefficient between the image in a square window **W$_n$** with side length *m* and starting point (1, 1) and the image in another window **w$_n$** whose starting point is ($x_0$, $y_0$) can be calculated as:

$$\Phi(1,1,x_0,y_0) = \frac{\sum_{p=0}^{m-x_0}\sum_{q=0}^{m-y_0}[W_n(1+p,1+q) \times w_n(x_0+p,y_0+q)]}{\{\sqrt{\sum_{p=0}^{m-x_0}\sum_{q=0}^{m-y_0}[W_n(1+p,1+q)]^2}\} \times \{\sqrt{\sum_{p=0}^{m-x_0}\sum_{q=0}^{m-y_0}[w_n(x_0+p,y_0+q)]^2}\}} \quad (B.1)$$

The displacement between the adjacent window **w$_n$** is set to be 0.5 time of the side length *m* to acquire the high precision results. In addition, in order to simplify the calculation, the cross-correlation analysis algorithm based on fast Fourier transform method is applied. Here we set:

$$\begin{cases} \mathbf{F_{W_n}} = FFT\{\mathbf{W_n}\} \\ \mathbf{F_{w_n}} = FFT\{\mathbf{w_n}\} \end{cases} \quad (B.2)$$

Then we have:

$$\begin{cases} f_n(i,j) = F_{Wn}(i,j)F_{wn}^*(i,j), \; 1 \leq i,j \leq m, \; i,j \in Z \\ \Phi = \frac{IFFT\{\mathbf{f_n}\}}{\sqrt{\sum_{p=1}^{m}\sum_{q=1}^{m}[W_n(p,q)]^2} \times \sqrt{\sum_{p=1}^{m}\sum_{q=1}^{m}[w_n(p,q)]^2}} \end{cases} \quad (B.3)$$

where $\Phi(i, j)$ is the correlation coefficient when the coordinate of the starting point of window **w$_n$** is (*i*, *j*). The coordinate of the point where **Φ** gets its maximum value (denoted as ($x_{max}$, $y_{max}$)) is the position of the starting point of the matching scheme with the largest correlation. Moreover, the value of $\Phi(x_{max}, y_{max})$ is denoted as $\Phi_n$, which means the correlation coefficient between images in two matched square windows **W$_n$** and **w$_n$**.

*Appendix C. The Derivation of Φ$_n$*

Since Equation (B.1) in Appendix B can be abbreviated as Equation (C.1), where $Vol_1(x_1+p, y_1+q)$ and $Vol_2(x_2+p, y_2+q)$ denote the volume of the liquid film in pixel ($x_1+p$, $y_1+q$) on square window **W$_n$** of figure $D_1$ and that in pixel ($x_2+p$, $y_2+q$) on square window **w$_n$** of figure $D_2$, respectively. With the application of the basic inequality, we can get Equation (C.2).

The derivation in Appendix C shows that $\Phi_n$ is related to the volume differences of the liquid film at the corresponding pixels of the two photos. Meanwhile, Equation (C.2) also shows that the differences on the volume of the liquid film in square windows **W$_n$** and **w$_n$** denotes the lower limit of $\Phi_n$. Combined with the Equation (12), we can see that $\Phi_n$ is related to the velocity of the liquid film. A higher $\Phi_n$ means a larger percentage of the liquid film whose velocity is equal to *v*.

$$\begin{aligned}\Phi_n(x_1,y_1,x_2,y_2) &= \frac{\sum_{p=0}^{m-x_1}\sum_{q=0}^{m-y_1}[(W_n(x_1+p,y_1+q) \times S_p) \times (w_n(x_2+p,y_2+q) \times S_p)]}{\{\sqrt{\sum_{p=0}^{m-x_1}\sum_{q=0}^{m-y_1}[W_n(x_1+p,y_1+q) \times S_p]^2}\} \times \{\sqrt{\sum_{p=0}^{m-x_1}\sum_{q=0}^{m-y_1}[w_n(x_2+p,y_2+q) \times S_p]^2}\}}\\ &= \frac{\sum_{p=0}^{m-x_1}\sum_{q=0}^{m-y_1}[(Vol_1(x_1+p,y_1+q)) \times (Vol_2(x_2+p,y_2+q))]}{\{\sqrt{\sum_{p=0}^{m-x_1}\sum_{q=0}^{m-y_1}[Vol_1(x_1+p,y_1+q)]^2}\} \times \{\sqrt{\sum_{p=0}^{m-x_1}\sum_{q=0}^{m-y_1}[Vol_2(x_2+p,y_2+q)]^2}\}} \end{aligned} \quad (C.1)$$

$$\Phi_n(x_1,y_1,x_2,y_2) \geq \frac{2\sum_{p=0}^{m-x_1}\sum_{q=0}^{m-y_1}[(Vol_1(x_1+p,y_1+q)) \times (Vol_2(x_2+p,y_2+q))]}{\sum_{p=0}^{m-x_1}\sum_{q=0}^{m-y_1}\{[Vol_1(x_1+p,y_1+q)]^2 + [Vol_2(x_2+p,y_2+q)]^2\}}$$

$$= 1 - \frac{\sum_{p=0}^{m-x_1}\sum_{q=0}^{m-y_1}[Vol_1(x_1+p,y_1+q)-Vol_2(x_2+p,y_2+q)]^2}{\sum_{p=0}^{m-x_0}\sum_{q=0}^{m-y_0}\{[Vol_1(x_1+p,y_1+q)]^2+[Vol_2(x_2+p,y_2+q)]^2\}} \quad (C.2)$$

*Appendix D. Symbol table*

The symbols in this work are listed in Table D1.

**Table D1.** Symbol table

| Parameter | Definition | Parameter | Definition |
|---|---|---|---|
| MT | the upper section of the main pipe | $r$ | the radios of the tee junction |
| MB | the lower section of the main pipe | $Vol(t_n)$ | the volume of liquid film at time $t_n$ |
| BR | the branch pipe | $\mathbf{W_n}$ | a square window in figure $D_1$ |
| $Q$ | the mass flow rate of the liquid film | $\mathbf{w_n}$ | a square window in figure $D_2$ |
| $\rho$ | the mass density of the liquid film | $\Phi_n$ | correlation coefficient between images in two matched windows $\mathbf{W_n}$ and $\mathbf{w_n}$ |
| $\bar{v}$ | the average flow velocity of the liquid film | $\bar{d}$ | the average thickness of the liquid film |
| $\bar{S}$ | the average cross-sectional vector area of the liquid film | $d_q(i)$ | the thickness distribution on the cross-section |
| $k$ | proportional coefficient | $\varepsilon$ | percentage error of the calculation results on liquid film's mass flow rate |
| $I_{ref}$ | the reference luminance | $q$ | the true value of liquid film's mass flow rate |
| $I\{n\}$ | the transmission of light under a certain wavelength | $m$ | the side length of square windows in the pictures |
| $I_{ref}\{n\}$ | the luminance of light under each wavelength contained in the filled light | $u(t)$ | the changing rate of liquid film's volume per unit length of the tee junction at time $t$ |
| $I$ | the brightness of the corresponding pixel in the subsequent image | | |